\pdfoutput=1
\documentclass[prb,showpacs,superscriptaddress,amsmath,amssymb]{revtex4}

\usepackage[latin1]{inputenc}

\usepackage[pdftex]{graphicx}
\usepackage{subfigure}
\usepackage[bookmarks]{hyperref}

\newcommand{\refsect}[1]{Sec.~\ref{#1}}
\newcommand{\reffig}[1]{Fig.~\ref{#1}}
\newcommand{\reffigpart}[2]{Fig.~\ref{#1}#2}
\newcommand{\refeq}[1]{Eq.~(\ref{#1})}
\newcommand{\reftab}[1]{Tab.~\ref{#1}}

\newcommand{\x}{X}
\newcommand{\y}{Y}
\newcommand{\z}{Z}

\newcommand{\VectorIII}[3]{
\begin{pmatrix} {#1} \\
                {#2} \\
                {#3}    
\end{pmatrix}
}

\newcommand{\MatFys}{
  Division of Mathematical Physics, Lund University, Box 118, S-22100 Lund, Sweden}

\newcommand{\SolidState}{
  Division of Solid State Physics, Lund University, Box 118, S-22100 Lund, Sweden}

\newcommand{\Duisburg}{
  Fachbereich Physik, Universität Duisburg-Essen, Germany}

\begin{document}

\title{\vspace{-0.5cm}Strain in Semiconductor Core-Shell Nanowires}

\author{Johan Grönqvist}
\affiliation{\MatFys}
\author{Niels Søndergaard \footnote{\vspace{-0.1cm} niels.sondergaard@matfys.lth.se}}
\affiliation{\MatFys}
\author{Fredrik Boxberg}
\affiliation{\SolidState}
\author{Thomas Guhr}
\affiliation{\MatFys}
\affiliation{\Duisburg}
\author{Sven Åberg}
\affiliation{\MatFys}
\author{H.Q.~Xu \footnote{\vspace{0.5cm} hongqi.xu@ftf.lth.se}}
\affiliation{\SolidState}


\begin{abstract}
  We compute strain distributions in core-shell nanowires of zinc
  blende structure. We use both continuum elasticity theory and an
  atomistic model, and consider both finite and infinite wires.  The
  atomistic valence force-field (VFF) model has only few
  assumptions. But it is less computationally efficient than the
  finite-element (FEM) continuum elasticity model. The generic
  properties of the strain distributions in core-shell nanowires
  obtained based on the two models agree well. This agreement
  indicates that although the calculations based on the VFF model are
  computationally feasible in many cases, the continuum elasticity
  theory suffices to describe the strain distributions in large
  core-shell nanowire structures.  We find that the obtained strain
  distributions for infinite wires are excellent approximations to the
  strain distributions in finite wires, except in the regions close to
  the ends. Thus, our most computationally efficient model, the
  finite-element continuum elasticity model developed for infinite
  wires, is sufficient, unless edge effects are important. We give a
  comprehensive discussion of strain profiles.  We find that the
  hydrostatic strain in the core is dominated by the axial
  strain-component, $\varepsilon_{\z \z}$. We also find that although
  the individual strain components have a complex structure, the
  hydrostatic strain shows a much simpler structure. All in-plane
  strain components are of similar magnitude. The non-planar
  off-diagonal strain-components ($\varepsilon_{\x \z}$ and
  $\varepsilon_{\y \z}$) are small but nonvanishing. Thus the material
  is not only stretched and compressed but also warped.  The models
  used can be extended for study of wurtzite nanowire structures, as
  well as nanowires with multiple shells.

\end{abstract}


\maketitle

\section{\label{sec:intro}Introduction}

Nanowires represent a promising technological platform with a wide
range of applications spanning from electronic \cite{HDCLKL01} and
photonic devices\cite{LGJSY05,WCZZYPM08,HDL05} to biochemical
sensors\cite{ZPCWL05}.  The reasons for this success are not only the
increasing ability in handling and manipulating nanowires but also an
ongoing improvement of the quality of the crystal structure of the
grown wires.

The progress in crystal growth has led to the fabrication of coherent
crystalline nanowires with core-shell structure\cite{Lauhon02}.  The
differences between the materials of the core and shell in lattice
constants and band parameters allow for tuning the properties of the
resulting wire \cite{Skold05}.  Possible strategies involve confining
charge carriers to the nanowire core to reduce the influence of the
surface on the electrical properties.  Likewise, a core-shell nanowire
can serve as an optical waveguide confining light. Confining charge
carriers to the shell, or having multiple active shells and confining
electrons and holes to different shells \cite{Schrier07, Zhang07}, may
also be relevant, depending on applications.  The confinement can be
obtained by band gap engineering and strain-engineering
\cite{Boxberg07}. In strain-engineering, one chooses to grow the core
and shell in a nanowire with different materials with a coherent
crystalline core-shell interface. Thus the nanowire is under a
\emph{pseudomorphic} strain, due to the deformations at the interface
to accommodate the different lattice constants.  This strain generates
different deformation potentials in the core and shell regions which
affect the electronic structure and thus the electrical and optical
properties of the nanowires.  In particular, gaining deep insight into
the electronic structure of heterostructured nanowires with a lattice
mismatch, requires finding the underlying elastic deformation, since
this elastic information is needed as input to, e.g., $k \cdot p $ or
tight-binding
calculations.\cite{Csontos08,Csontos09,Persson02,Persson04PRB,Persson04Nano,Persson06PRB1,Persson06PRB2}

Theoretical investigations based on the continuum elasticity of
systems experiencing pseudomorphic strain go back to Eshelby
\cite{Eshelby} who used analytical methods to study the strain effects
of one material immersed in another. More recently, several studies of
pseudomorphic strain fields have been performed for finite
nanostructures.\cite{Stangl04} On core-shell nanowire geometries,
Niquet studied the effect of a shell on the electronic properties of a
quantum dot embedded in a wire.\cite{Niquet06, Niquet07} Studies on
the effect of a core-shell geometry on the electronic properties in
quantum wires were also performed by Pistol and Pryor, \cite{Pistol08}
Schrier {\it et al},\cite{Schrier07} and Zhang {\it et
  al}.\cite{Zhang07} However, these studies only considered very small
structures or presented the hydrostatic strain in the structures.

In this article, we calculate and discuss the elastic deformation in
core-shell nanowires of zinc blende structure with a lattice mismatch
between the core and shell materials.  Numerical calculations are
performed for both finite and infinite wires using both continuum
elasticity theory (implemented by a finite element scheme) and the
atomistic model of Keating \cite{Keating66}.  We present the strain
distributions in a cross section of the nanowire as well as line scans
both in the cross-section and along the wire.  We consider free
nanowires, i.e., we neglect external forces acting on the nanowires.
Core-shell nanowires with various cross sections are investigated:
hexagonal cross sections with both parallel and non-parallel core and
shell facets and co-centric circular cross sections.  We restrict
ourselves to only discussing nanowires with a single shell and a core,
although the methodology used here can be generalized to multiple
shell nanowires. Likewise, we shall only consider purely zinc blende
crystalline heterostructures.

The article is organized as follows. The finite-element continuum
elasticity model employed in this work is described in
\refsect{sec:CE} and the atomistic model in
\refsect{sec:atomistic}. Numerical results and discussion are
presented in \refsect{sec:data}. Finally, \refsect{sec:conclusions} is
devoted to summary and conclusions.

\section{\label{sec:CE} Continuum elasticity theory}

We here describe the continuum elasticity (CE) model employed for a core-shell nanowire
of zinc blende structure with axis along the $\left[111\right]$ direction
 for which the material lattice parameters vary discontinuously at the core-shell interface.

\subsection{Coordinate systems}
\label{sec:Geometry}

\begin{figure}[ht]
\begin{center}
\includegraphics[width=8cm]{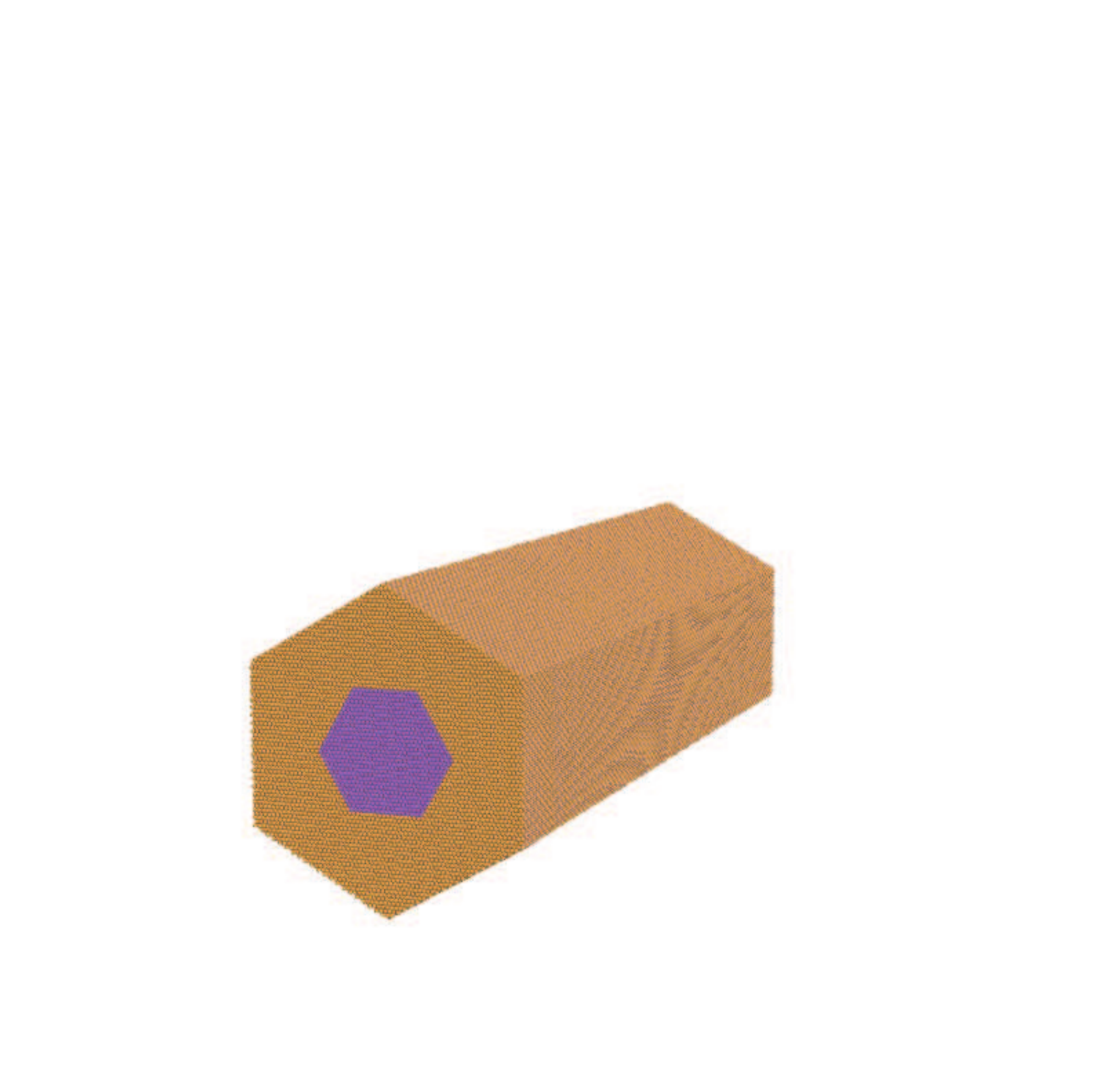}
\caption{\label{fig:3dWire}Nanowire with core and shell. Our results
  refer to a core of GaAs inside a shell of GaP. The core hexagon has
  a side-length of 6 nm and the shell hexagon has a side-length of
  13.9 nm.  The axis direction is $[111]$.}
\end{center}
\end{figure}

\begin{figure}[t] 
\includegraphics[width=5cm]{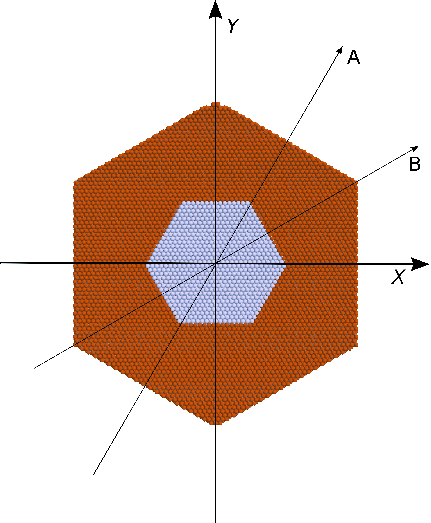}
\caption{\label{fig:CrossSec}Cross-section view of the core-shell geometry. Coordinate
  axes $X$ and $Y$ indicate our nanowire oriented coordinate system. The indicated paths A and B  are
  used when we present our numerical results in \refsect{sec:data}.}
\end{figure}

We consider core-shell nanowires of zinc blende structure with axis along  the
$\left[111\right]$-direction. Figure \ref{fig:3dWire} shows a typical
core-shell nanowire geometry. We define two coordinate systems: a standard
coordinate system with axes denoted by $(x, y, z)$ coinciding with the
main crystallographic axes $([100], [010], [001])$ and another system
where the axes will be denoted by $(\x, \y, \z)$, referred to as the
nanowire-oriented coordinate system. In the $(\x, \y, \z)$ system, the $\z$ axis is
parallel to the axis direction of the nanowire (i.e., the $[111]$
direction). The $\x$ and $\y$ axes are in the $[1 \bar{1} 0]$
and $[11 \bar{2}]$ crystal directions, respectively, as 
depicted in \reffig{fig:CrossSec}. Formally, the nanowire oriented
coordinate system is related to the standard coordinate system by the transformation
\begin{equation}
  \VectorIII{\x}{\y}{\z} = \mathcal{R} \VectorIII{x}{y}{z},
\end{equation}
with a rotation matrix $\mathcal{R}$ given by
\begin{equation}
\mathcal{R}=
\begin{pmatrix}
\frac{1}{\sqrt{2}} & -\frac{1}{\sqrt{2}} & 0 \\
\frac{1}{\sqrt{6}} & \frac{1}{\sqrt{6}} & -\frac{2}{\sqrt{6}}   \\
\frac{1}{\sqrt{3}} & \frac{1}{\sqrt{3}} & \frac{1}{\sqrt{3}}  \\
\end{pmatrix} \,. 
\end{equation}

\subsection{Strain energy}
\label{sec:StrainEnergy}

The strain energy $U$ is written as
\begin{equation}
\label{eq:StrainEnergy}
U = \int w \, dV \equiv \frac{1}{2} \, \int  \sum_{ijkl} c_{i j k l} \, \varepsilon_{i j}\, \varepsilon_{k l} \, dV  \,,
\end{equation}
where $w$ is the strain energy density, $c_{ijkl}$ are the elastic
stiffness tensor elements, and $\varepsilon_{ij}$ are strain tensor components
\cite{Nye85}.

For a material with cubic symmetry, the elastic stiffness tensor, $c_{ijkl}$, only has three independent
elements and the strain energy density in the crystallographic coordinate
system can be written as 
\begin{equation}
w = \frac{1}{2} c_{11} (\varepsilon_{xx}^2 +
\varepsilon_{yy}^2 + \varepsilon_{zz}^2) + c_{12} (\varepsilon_{xx}
\varepsilon_{yy} + \varepsilon_{xx} \varepsilon_{zz} +
\varepsilon_{yy} \varepsilon_{zz}) + 2 c_{44} (\varepsilon_{xy}^2 +
\varepsilon_{xz}^2 + \varepsilon_{yz}^2),
\end{equation}
where
\begin{eqnarray} \label{def:Constants}
  c_{11} = c_{iiii} &  &(i = x, y, z) \, , \nonumber \\
  c_{12} = c_{iijj} &  & (i, j = x, y, z; \, i\ne j) \, , \nonumber \\
  c_{44} = c_{ijij} &= c_{ijji} \; \; & (i, j = x, y, z;\,  i\ne j) \,.
\end{eqnarray}
The  identities in \refeq{def:Constants} reflect the cubic
symmetry of the material and define the general symmetry properties of the elasticity tensor.\cite{Nye85} In our calculations,  numerical values of the $c_{ij}$ are taken from Ref.~\onlinecite{Vurgaftman2001}.

\subsection{Incorporating pseudomorphic strain }
\label{sec:TwoMaterial}
In this section we describe the lattice matching conditions and their
interpretation in two- and three-dimensional bulk systems.

\subsubsection{Pseudomorphic conditions}
\label{sec:PseudoMorph}
We consider two different elastic media, which are the core and the
shell materials of the same crystal structure. We shall use 
superscripts $(c)$ and $(s)$ to indicate core and shell, respectively.  A
typical geometry for a finite wire is shown in \reffig{fig:3dWire},
and a cross section of the wire is shown in \reffig{fig:CrossSec}.

For cubic materials the lattice vectors $\mathbf{a}^{(c)}$ and
$\mathbf{a}^{(s)}$ of core and shell in the strain-free state can be
written as
\begin{eqnarray}
\mathbf{a}^{(c)} = \sum_i a^{(c)}_i \mathbf{e}_i=a^{(c)} \sum_i n_i \mathbf{e}_i ,\nonumber\\
\mathbf{a}^{(s)} = \sum_i a^{(s)}_i \mathbf{e}_i=a^{(s)} \sum_i n_i \mathbf{e}_i,
\end{eqnarray}
where $a^{(c)}$ and $a^{(s)}$ are the lattice constants of the core
and shell materials, $a^{(c)}_i$ and $a^{(s)}_i$ are the components of
the lattice vectors in the core and shell, $n_i$ are integers, and
$\mathbf{e}_i$ are the basic vector in the crystallographic coordinate
system.

After lattice deformation, distorted lattice vectors in the core and
shell can be written in terms of the components $\varepsilon^{(c)}_{i
  j}$ and $\varepsilon^{(c)}_{i j}$ of the corresponding strain
tensor, as
\begin{eqnarray}
\mathbf{r}^{(c)} = \sum_i \mathbf{e}_i\sum_j \left(\delta_{i j} + \varepsilon^{(c)}_{i j} \right) a^{(c)}_j ,\nonumber\\
\mathbf{r}^{(s)} = \sum_i \mathbf{e}_i\sum_j \left(\delta_{i j} + \varepsilon^{(s)}_{i j} \right) a^{(s)}_j .
\end{eqnarray}
At the core-shell interface, the in-plane component of the distorted
vector $\mathbf{r}^{(c)}$ on the core side must match its equivalent
$\mathbf{r}^{(s)}$ on the shell side. This pseudomorphic requirement
can be expressed as
\begin{equation}
\label{eq:TensPseudomCond}
\sum_{ij} t_i \left[\delta_{i j} + \varepsilon^{(c)}_{i j} \right] a^{(c)}_j =\sum_{ij} t_i \left[\delta_{i j} + \varepsilon^{(s)}_{i j} \right] a^{(s)}_j, 
\end{equation}
where $t_i$ are the components of an arbitrary tangent vector $\mathbf{t}$ of the interface.
One implication of \refeq{eq:TensPseudomCond} is that the strain will
be discontinuous at an interface between two cubic materials having
different lattice constants.

\subsubsection{Initial strain}
\label{sec:InitialStrain}

To fulfill the conditions of \refeq{eq:TensPseudomCond}, we assume the
strain tensor in \refeq{eq:StrainEnergy} is given by  a sum of two terms
\begin{equation}
  \label{eq:CE_dof}
\boldsymbol{\varepsilon} =\frac{1}{2}\left[ \boldsymbol{\nabla}\otimes \mathbf{u}+ (\boldsymbol{\nabla}\otimes \mathbf{u})^t \right] +\boldsymbol{\varepsilon}^{(0)}\, ,
\end{equation} 
where $\mathbf{u}$ is the displacement field relative to a matched (yet unrelaxed) configuration and 
$\boldsymbol{\varepsilon}^{(0)}$ is an initial strain\cite{PdC06}.

The initial strain  is assumed nonvanishing only in the shell, where it is defined, in the cubic case, by  
\begin{equation} 
  \boldsymbol{\varepsilon}^{(0)} = \frac{a^{(c)}- a^{(s)}}{a^{(s)}} \cdot \mathbf{1} \,
\end{equation}
where $a^{(c)}$ and $a^{(s)}$ are the lengths of the lattice vectors
in the core and the shell, respectively. This initial strain ensures
that the conditions \refeq{eq:TensPseudomCond} are satisfied and leads
to a matched structure.

The structure obtained for $\mathbf{u}=0$ is not that of lowest energy
for $\boldsymbol{\varepsilon}^{(0)} \neq 0$ and the structure is
relaxed by varying the field $\mathbf{u}$ to minimize the energy
\refeq{eq:StrainEnergy}. For a finite wire, the degrees of freedom to
vary are precisely the degrees of freedom of $\mathbf{u}$. For the
case of an infinite wire, the periodicity in the $\z$-direction is
utilized and the resulting model is described below.

\subsection{Infinite Wire}
\label{sec:WireCase}

\label{sec:TransformEandU}
The model of an infinite wire utilizes the periodicity in the
$\z$-direction and reduces the modeling domain to a 2-dimensional domain.

The strain tensor is  transformed to the nanowire oriented coordinate system by 
\begin{equation}
\varepsilon_{\alpha \beta} = \sum_{a b}\mathcal{R}_{\alpha a} \mathcal{R}_{\beta b} \, \varepsilon_{a b} \,.
\end{equation}
where Greek indices vary over $\x,$ $\y,$ and  $\z$ and Latin indices  vary over $x,$ $y,$ and $z$.

In the model of an infinite wire, the vectorial field $\mathbf{u}$ introduced in
\refeq{eq:CE_dof} depends only on the coordinates in an $(\x, \y)$
plane of the nanowire, and we write $\mathbf{u}(\x, \y) \in
\mathbb{R}^3$.  The axial strain in the core $\varepsilon_{\z \z}^{(c)}$ and the axial strain in the shell $\varepsilon_{\z \z}^{(s)}$ are constant and   by \refeq{eq:TensPseudomCond} fulfill
\begin{equation}
  \label{eq:AxialCond}
  \left[1+\varepsilon^{(c)}_{\z \z} \right]\, a^{(c)} = \left[1+\varepsilon^{(s)}_{\z \z}\right]\, a^{(s)} \,.
\end{equation}
The axial strain can thus be described as a single
degree of freedom $a$ as 
\begin{equation}
\label{eq:AxStrainAndA}
  \varepsilon_{\z \z}^{(i)} = \frac{a}{a^{(i)}} - 1 \qquad \text{with} \qquad i=s \text{ or } c \,.
\end{equation}
In the model of an infinite wire, the field $\mathbf{u}(\x, \y)$ and the  variable $a$ are the degrees of freedom.

The energy density [defined in \refeq{eq:StrainEnergy}] becomes
\begin{eqnarray}
\label{eq:UnstrainedEnergy}
w &=& \frac{1}{2} \left[ F_1 (\varepsilon_{\x \x }^2 + \varepsilon_{\y \y }^2)+ F_2 \varepsilon_{\x \x } \varepsilon_{\y \y } + F_3 (\varepsilon_{\x \x } + \varepsilon_{\y \y }) \varepsilon_{\z \z } \right.  \nonumber \\ 
\nonumber & &  \left. + F_4 (\varepsilon_{\x \x }-\varepsilon_{\y \y })\varepsilon_{\y \z }    + 2 F_4 \varepsilon_{\x \z } \varepsilon_{\x \y }+F_5 \varepsilon_{\z \z }^2  \right.  \\  
& &  \left. +F_6 (\varepsilon_{\y \z }^2 + \varepsilon_{\x \z }^2)+F_7 \varepsilon_{\x \y }^2 \right] \,
\end{eqnarray}
with the constants
\begin{eqnarray}
F_1 &=& \frac{1}{2} \left(c_{11}+c_{12}+2 c_{44}\right)  , \\ \nonumber
F_2 &=& \frac{1}{3}  \left(c_{11}+5 c_{12}-2 c_{44}\right) , \\ \nonumber
F_3 &=&   \frac{2}{3}   \left(c_{11}+2 c_{12}-2 c_{44}\right) , \\ \nonumber
F_4 &=& \frac{2\sqrt{2}}{3} \left( c_{11} -c_{12} -2 c_{44}  \right) , \\ \nonumber
F_5 &=& \frac{1}{3}  \left(c_{11}+2 c_{12}+4 c_{44}\right)  , \\ \nonumber
F_6 &=& \frac{4}{3} \left( c_{11} -c_{12} + c_{44}  \right)  , \\ \nonumber
F_7 &=&   \frac{2}{3}   \left(c_{11}-c_{12}+4 c_{44}\right) \, .
\end{eqnarray}

In this model, we keep  $\varepsilon_{\x \z}$ and
$\varepsilon_{\y \z}$ in contrast to  the conventional plane strain approximation  in  which $\varepsilon_{\x \z}=\varepsilon_{\y \z}=0$ is assumed\cite{Cl03,SHFHGX09}.

The minimization of the strain energy over the  variable $a$ corresponds to the
condition that the total axial force $F_\z$ vanishes. This has been
discussed for a simpler energy functional in Ref.~\onlinecite{SHFHGX09}.  The vanishing of the total axial force can be understood
by considering a wire of a fixed number of unit cells
$N$. By translation invariance the total energy is proportional to
$N$. The variation of the energy per unit cell becomes
\begin{eqnarray}
0 &=& \frac{1}{N} \, \frac{\partial U}{\partial a} \\ \nonumber
 &=&  \frac{1}{N} \, \bigglb( \frac{\partial U_{\mathrm{c}}}{\partial a}+ \frac{\partial U_{\mathrm{s}}}{\partial a} \biggrb)
\end{eqnarray}
where $U_{\mathrm{c}}$ and  $U_{\mathrm{s}}$ are the strain energies of core and shell.
The energy is given by Eq.~(\ref{eq:StrainEnergy}) with the integration
taken over the undeformed domains \cite{PdC06}. For the core the
undeformed axial length $l_{\mathrm{c}}$ is proportional to $N
a^{(\mathrm{c})}$ and for the shell the axial length 
$l_{\mathrm{s}}$ is proportional to $N a^{(\mathrm{s})}$.
In particular, the total energy of the core with energy density $w_{\mathrm{c}}$ is given by
\begin{equation}
U_\mathrm{c} = \int_{0}^{l_{\mathrm{c}}} dz \int_{\mathrm{c}} dS \, w_{\mathrm{c}} \propto N a^{(\mathrm{c})} \int_{\mathrm{c}} dS \, w_{\mathrm{c}} \,.
\end{equation}
For the variation of the core energy, we calculate the partial derivative
\begin{equation}
\frac{\partial w_{\mathrm{c}}}{\partial a} = \frac{1}{a^{(\mathrm{c})}} \frac{\partial w_{\mathrm{c}}}{\partial \varepsilon_{\z \z}} =  \frac{ \sigma_{\z \z}}{a^{(\mathrm{c})}}  \,,
\end{equation}
where $\sigma_{\z \z} = F_5 \varepsilon_{\z \z} + F_3 (\varepsilon_{\x \x}+\varepsilon_{\y \y})/2$ is the axial stress. The  area element
$dS$  and the axial stress then correspond to
an infinitesimal axial force $dF_{z}= dS \,  \sigma_{\z \z}$.  
Using these results and those similar for the shell, we can write 
\begin{eqnarray}
\label{eq:VanishForce}
0 &=& \frac{1}{N} \, \frac{\partial U}{\partial a} \\ \nonumber
&\propto&  \int_{\mathrm{c}} dS \,  \sigma_{\z \z}+ \int_{\mathrm{s}} dS \,  \sigma_{\z \z} \equiv  \,  \int_{\mathrm{c}} dF_\z + \int_{\mathrm{s}} dF_\z = F_{\z} \,
\end{eqnarray}
proving the assertion. This result [Eq.~(\ref{eq:VanishForce})] could be generalized to infinite wires
under stretch. In that case we would use a  non-zero force condition.

The actual minimization is done using finite elements \cite{zienkiewicz-89} (linear triangular elements) for the continuous
displacement $\mathbf{u}$ and a  variable $a$ to parameterize the
axial strain [\refeq{eq:AxStrainAndA}]. The finite element method (FEM) is
chosen due to its flexibility with respect to arbitrary
geometries. The method leads to a sparse matrix problem of the form $\mathbf{F} = \mathbf{K \, x }$ from which $\mathbf{x}$ (representing $\mathbf{u}$ and $a$) is solved by standard numerical
packages. Note that here the  variable $a$ couples to all other degrees
of freedom in the matrix $\mathbf{K}$. This is unlike the usual setting of only local coupling
in  two- and three dimensional bulk structures. Still, in practice the
solution is very fast even for meshes with around 30000 nodes and three
degrees of freedom per node.

\subsection{Finite Wire}
\label{sec:FiniteWire}
\label{sec:saint_venant}

We also compute the strain in finite core-shell nanowires. The
strain is computed by minimizing the total strain energy, defined in
Eq. \ref{eq:StrainEnergy}. This is performed in the coordinate
system $x$, $y$, and $z$ of the principal axes of the underlying
crystal and by incorporating the pseudomorphic strain as described in
\refsect{sec:TwoMaterial}. For finite wires we have no benefit of using the
nanowire oriented coordinate system because of the lack of
translational symmetry.

The typical length $L_z$ of a modeled finite wire range from $30\,$nm
to $ 150\,$nm. In the modeling of finite wires we exploit the
$C_{3v}$ symmetry of the nanowire. That is, we model and mesh only
a segment of $1/6$ of the total nanowire geometry. On the nodes on the
side facets of this segment we then fix the displacement
perpendicular to the facet\cite{footFBSymm}. By this
symmetry reduction we reduce the computational domain and the number
of nodes by a factor of $6$. This technique also reduce the source of
numerical errors since the resulting element mesh has the same
symmetry as the nanowire geometry.

The CE model of finite wires is implemented using
three-dimensional tetrahedron elements with second order
polynomials. A typical element mesh contain in total around $1.5
\cdot 10^5$ nodes distributed in a wire, with 3 degrees of freedom per
node. We use a nonuniform and adapted element mesh with at most $9
\cdot 10^4$ nodes per cross-sectional area (on $1/6$ of the nanowire
cross-section).

\section{\label{sec:atomistic}Atomistic VFF Model}

For the atomistic calculations we use the Valence-Force Field 
(VFF) model of Keating\cite{Keating66}. The structure is built up one atomic
layer at a time.  We choose a layer structure that corresponds to a
zinc blende phase nanowire with axis along the $\left[111\right]$
direction. 

We first describe our model for finite wires and then describe our
implementation for an infinite wire using translation invariance.

\subsection{Finite wires}

The atomic coordinates
$\{\mathbf{r}_i\}_{i=1}^N$ are the degrees of freedom of our VFF model. The energy is a function of these and is written as
sums over interatomic bond distances and three-body bond angles.

We let $\mathbf{r}_{ij} = \mathbf{r}_{j} - \mathbf{r}_{i}$ denote the vector from atom $i$ to
atom $j$ and $\mathbf{r}_{ij0}$ denote the value $\mathbf{r}_{ij}$ takes in a
stress-free bulk crystal. The potential energy of the crystal is
given by \cite{Keating66}
\begin{eqnarray}
  \label{eq:keating_energy}
  E &=&  \sum_i \sum_j^\mathrm{n.n.} \frac{3 \alpha}{8 \mathbf{r}_{ij0}^2} \left(\mathbf{r}_{ij}^2 - \mathbf{r}_{ij0}^2\right)^2  \\
  && + \sum_i \sum_{j, k\ne j}^{\mathrm{n.n.}} \frac{3 \beta}{8 \left|\mathbf{r}_{ij0}\right| \left|\mathbf{r}_{ik0}\right|} \left( \mathbf{r}_{ij} \cdot \mathbf{r}_{ik} - \mathbf{r}_{ij0} \cdot \mathbf{r}_{ik0} \right)^2 \nonumber
\end{eqnarray}
where the equilibrium bond angles fulfill $\mathbf{r}_{ij0} \cdot \mathbf{r}_{ik0} =
-\left|\mathbf{r}_{ij0}\right| \left|\mathbf{r}_{ik0}\right|/3$ in zinc blende
and n.n. on top of a summation sign indicates that we sum over nearest
neighbors of  atom $i$ only.

Numerical values for the coupling constants $\alpha$ and $\beta$ are
obtained by comparing the energy in the VFF model
\refeq{eq:keating_energy} and the CE model \refeq{eq:StrainEnergy},
and requiring that the energies coincide for small deformations of a small
volume. This requires three equations to be satisfied, one for each
material parameter in the CE model. As the VFF model only contains the
two coupling constants $\alpha$ and $\beta$, all three equations
cannot be exactly satisfied, but any choice of values for $\alpha$ and
$\beta$ will give rise to effective elastic constants of the VFF model
that deviate form the experimentally determined values used in the CE
model.  We follow Pryor et. al. \cite{Pryor98} in using the values
provided by Martin \cite{Martin70}. Numerical values for the coupling
constants and the relative errors in effective elastic constants are
summarized in \reftab{tab:vff_cpl_consts}.

\begin{table}[hp]
\centering
\caption{\label{tab:vff_cpl_consts} Coupling constants used in the VFF model, 
and relative errors for effective elastic constants.}

\begin{ruledtabular}
\begin{tabular}{lrr}

& GaAs & GaP \\
\hline
$\alpha$ ($ \mathrm{N/m}$) & 41.19 & 47.32 \\ 
$\beta$ ($ \mathrm{N/m}$) & 8.95 & 10.44 \\
$r_0$ (Å) & 2.448 & 2.36 \\
$\delta c_{11}$ (\%)& -0.22 & 2.7 \\
$\delta c_{12}$ (\%)& 1.7 & 9.1 \\
$\delta c_{44}$ (\%)& -13 & -11 \\

\end{tabular}
\end{ruledtabular}
\end{table}

Possible extensions of our model include an additional Coulomb
interaction energy \cite{Martin70} and higher order terms for
non-linear elasticity \cite{Keating66b, Cousins03}. Such extensions
are beyond the scope of the current work.  The current model is widely
applicable for analyzing linear elasticity of, e.g., zinc blende and
wurtzite crystals. It could also be applied to the description of
epitaxially strained heterointerfaces between materials with different
lattice structures (see e.g.~\onlinecite{WCZZYPM08}). It is, however,
worth to note that the applicability of the VFF model is, in this case,
dependent on a priori information about the interface structure.

\subsection{Infinite wires}

For the simulation of an infinite wire, we start with a finite wire
segment. We consider this finite segment to be a 
unit cell of an infinite wire  consisting of an infinite
array  of such  unit cells. The distance between two unit cells
in the axial direction is added to the model as an additional degree
of freedom $\Delta r$ and is called the axial lattice constant in the
following.

Bonds between the different unit cells are added to the energy
expression for a finite wire segment which was given in
\refeq{eq:keating_energy}. This gives an expression for the energy per
unit cell of an infinite wire.  The inter-cell bonds couple the
degrees of freedom of the topmost and bottommost atoms, via the axial
lattice constant $\Delta r$. We then minimize the energy per unit
cell, the degrees of freedom now being the atomic positions inside the
unit cell and the axial lattice constant $\Delta r$.

\subsection{Definition of strain}

To enable comparisons with
continuum elasticity we must define a deformation measure from the
atomistic VFF displacements that is comparable to the macroscopic
elastic strain.  Strain is a macroscopic concept 
and several different atomistic strain quantities can
be defined, that all converge to the macroscopic strain in the
relevant limit. In this work, we follow the strain definition of
Pryor and coworkers \cite{Pryor98}.

\begin{figure}
\includegraphics[width=6cm]{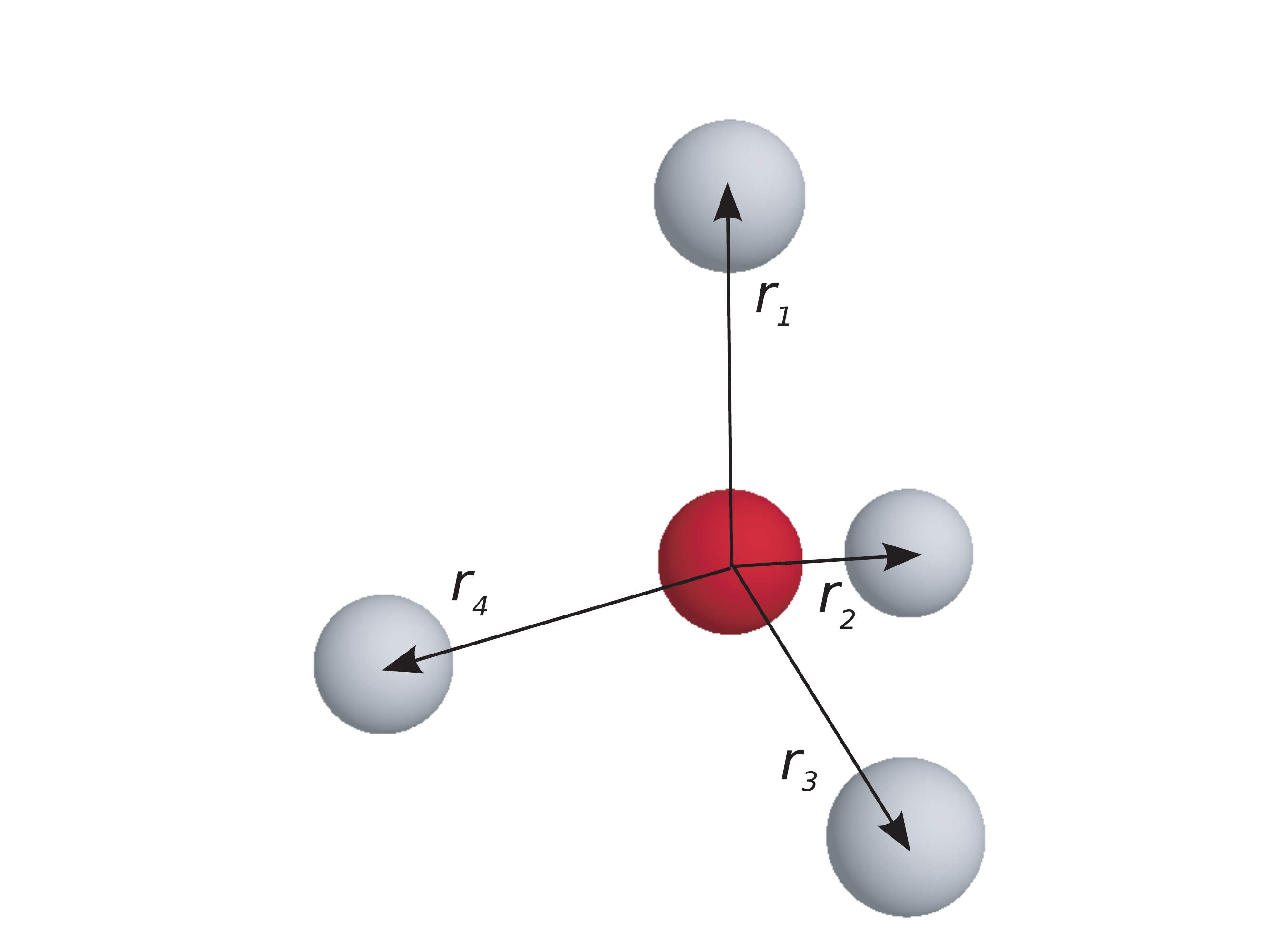}
\caption{\label{fig:tetrahedron}The tetrahedron of neighbors around one atom in the atomistic model.}
\end{figure}

\subsubsection{Microscopic definition}

The microscopic definition of strain at the position of an atom is formulated in terms of displacements of the atoms in its local neighborhood. 
For an atom, positioned at $\mathbf{r}_0$, the relative position
vectors $\mathbf{r}_1, \ldots, \mathbf{r}_4$ of its neighbors form a
tetrahedron around it as shown in \reffig{fig:tetrahedron}. In order
to define our concept of strain at $\mathbf{r}_0$, we consider these  
5 atoms to form the local neighborhood of the atom at $\mathbf{r}_0$. Our definition of strain
will use a deformed and an undeformed configuration of this neighborhood. 
In the deformed configuration, the positions of the 5 atoms
are the same as in the result of the full VFF calculation. In the other,
the undeformed configuration, the positions of the atoms are chosen
such that the energy of the isolated 5-atom system considered, as
defined by the VFF model \eqref{eq:keating_energy}, vanishes. In the
cases of a pure GaAs or a pure GaP tetrahedron, the undeformed
configuration is equal to a stress-free bulk configuration. At
interfaces, there is no corresponding bulk system.

For each of the two configurations, we define the vectors $\mathbf{r}_{ij} =
\mathbf{r}_j - \mathbf{r}_i$ from the coordinates of the tetrahedron corners, and from
these vectors we construct a $3 \times 3$ matrix whose columns are given
by $\mathbf{r}_{21}$, $\mathbf{r}_{32}$ and $\mathbf{r}_{43}$. Strain is then defined \cite{Pryor98} via
\begin{eqnarray}
  \label{eq:atomistic_strain}
  \left[
  \begin{array}{ccc}
    \mathbf{r}_{21}^{(d)} & \mathbf{r}_{32}^{(d)} & \mathbf{r}_{43}^{(d)} 
  \end{array}
  \right] &=& (\mathbf{1} + \tilde{\boldsymbol{\varepsilon}})
  \left[
  \begin{array}{ccc}
    \mathbf{r}_{21}^{(0)} & \mathbf{r}_{32}^{(0)} & \mathbf{r}_{43}^{(0)} 
  \end{array}
  \right] \nonumber \\
  \boldsymbol{\varepsilon} &=& \frac{\boldsymbol{\tilde{\varepsilon}} + \boldsymbol{\tilde{\varepsilon}}^T}{2}\,,
\end{eqnarray}
where the superscripts $d$ and $0$, refer to the coordinates of the deformed and undeformed state, respectively.
The strain at $\mathbf{r}_0$ is, consequently, the symmetric part of the tensor
that describes, to lowest order, the  deformation of a small volume around
the atom at $\mathbf{r}_0$. 

\subsubsection{Numerical average}

For the microscopic strain defined by (\ref{eq:atomistic_strain}), we
find strong oscillations in the strain field. Neighboring atoms can
have very different strains. This is particularly clear at the
\mbox{GaAs-GaP} interface, where the microscopic description of the interface
renders it non-planar (even locally), and this is phenomenologically
different from the continuum elastic description of the same
interface. In order to allow for a comparison of the strain, obtained
with these two elasticity models, we have chosen to smoothen the
strain of the VFF model. We perform this smoothing by averaging the
strain over each pair of group III and group V atoms aligned along the $[111]$ direction.

The static strain field of continuum elasticity describes macroscopic
features of the continuous material, and thus implicitly contains a smoothing
over any internal deformation at the atomic scale. The
strong variations we find on this scale can thus have no direct
correspondence in continuum elasticity, and our averaging procedure
removes those effects.

\section{\label{sec:data}Results}

When comparing the results for finite and infinite nanowires, we
evaluate the strain of a finite wire at cross sections in the
$XY$-plane far enough from the free ends of the nanowire. We only
discuss the strains in the $(\x, \y, \z)$-system.  By the principle of
Saint-Venant\cite{Cl03}, the actual force and moment distributions at
the cross-sections are not important for the state far enough from the
cross sections. Only the total force and moment matter.  For a
meaningful comparison, we therefore require that the resultant force
and moment on these ends are similar to the conditions on the cross
section of the infinite wire.  In our case, we consider free nanowires
without external forces. This corresponds to no total force on the
cross-sections.  As we expect, the results for a cross section of the
finite models converge to those of the infinite models with increasing
distance of the cross section from the free ends.

Deviations between the atomistic and continuum models are expected to
be dominated by the difference in effective elastic constants. The
force parameterization in the VFF model uses only two parameters that
are to be determined from the three material parameters of cubically
symmetric continuum elasticity. We find that this imperfect matching is
indeed the main source of discrepancies between the results of the VFF
and CE calculations.

\subsection{Strain distribution}

\begin{figure*}
  \includegraphics[width=\textwidth]{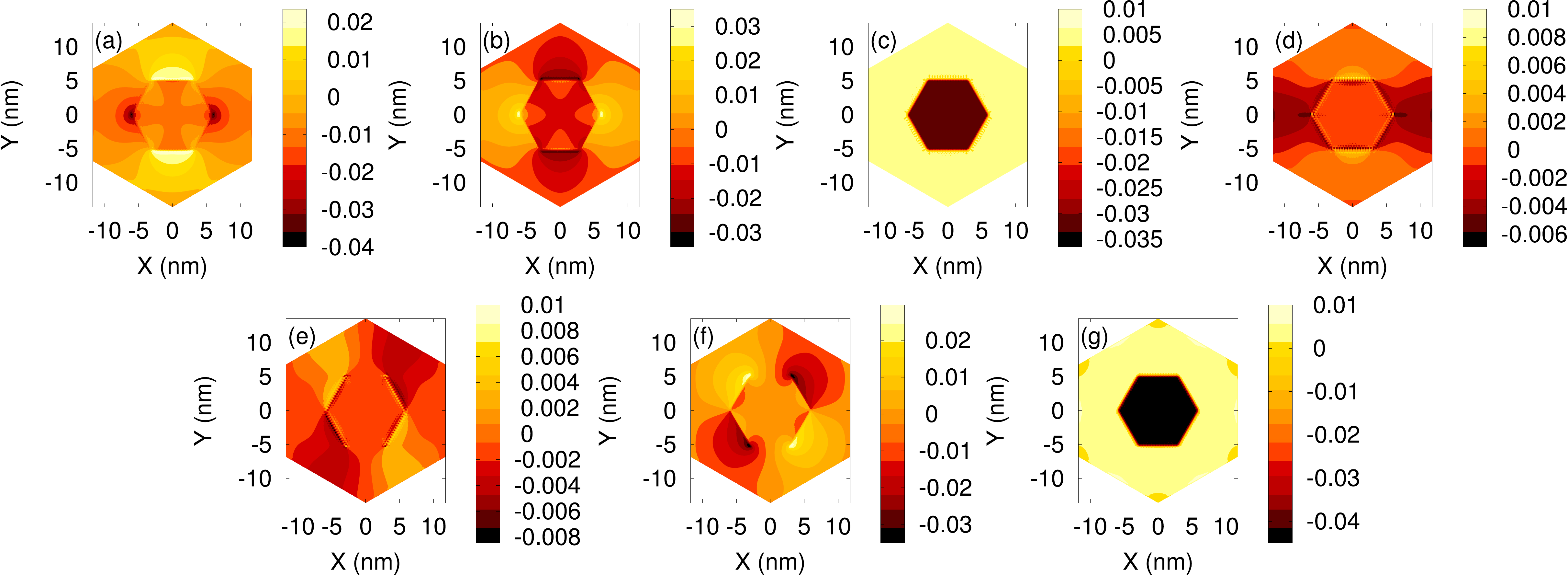}
  \caption{\label{fig:strain_plots_first}\label{fig:zb_strain}Strain in an axial unit cell of an
  infinite nanowire with a hexagonal geometry. The results were
  obtained with the infinite atomistic model. The plots (a) to (g)
  show the strain components $\varepsilon_{\x \x}$, $\varepsilon_{\y
    \y}$, $\varepsilon_{\z \z}$, $\varepsilon_{\y \z}$,
  $\varepsilon_{\x \z}$, and $\varepsilon_{\x \y}$ and the hydrostatic
  strain $\varepsilon_{\x \x} + \varepsilon_{\y \y} + \varepsilon_{\z
    \z}$, respectively. The axes are as indicated in
  \reffig{fig:CrossSec}.}
\end{figure*}

\begin{figure*}
  \includegraphics[width=\textwidth]{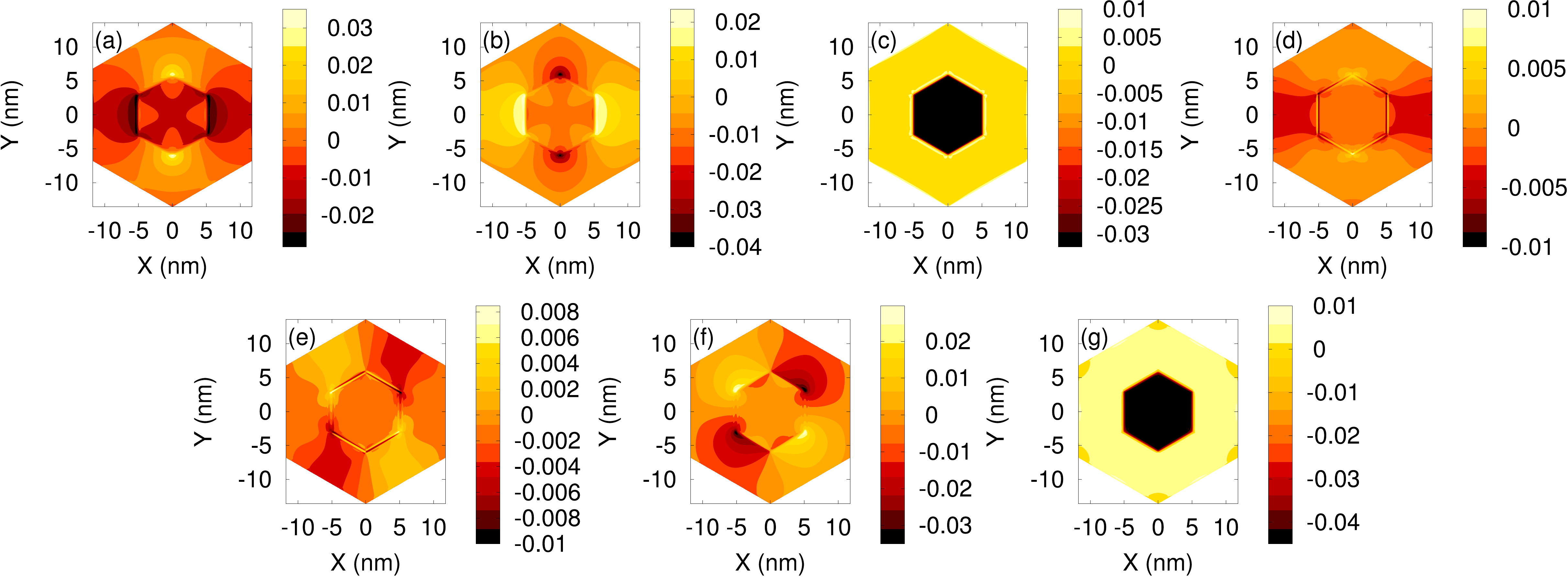}
  
  \caption{\label{fig:strain_plots_middle}\label{fig:zb_aligned_strain}Strain fields in a unit cell of an infinite nanowire where the
  core and shell hexagons have the same orientations. Results obtained
  from the infinite atomistic model. The plots (a)-(g) show the strain
  components $\varepsilon_{\x \x}$, $\varepsilon_{\y \y}$,
  $\varepsilon_{\z \z}$, $\varepsilon_{\y \z}$, $\varepsilon_{\x \z}$, and
  $\varepsilon_{\x \y}$ and the hydrostatic strain $\varepsilon_{\x
    \x} + \varepsilon_{\y \y} + \varepsilon_{\z \z}$, respectively.}
\end{figure*}

\begin{figure*}
  \includegraphics[width=\textwidth]{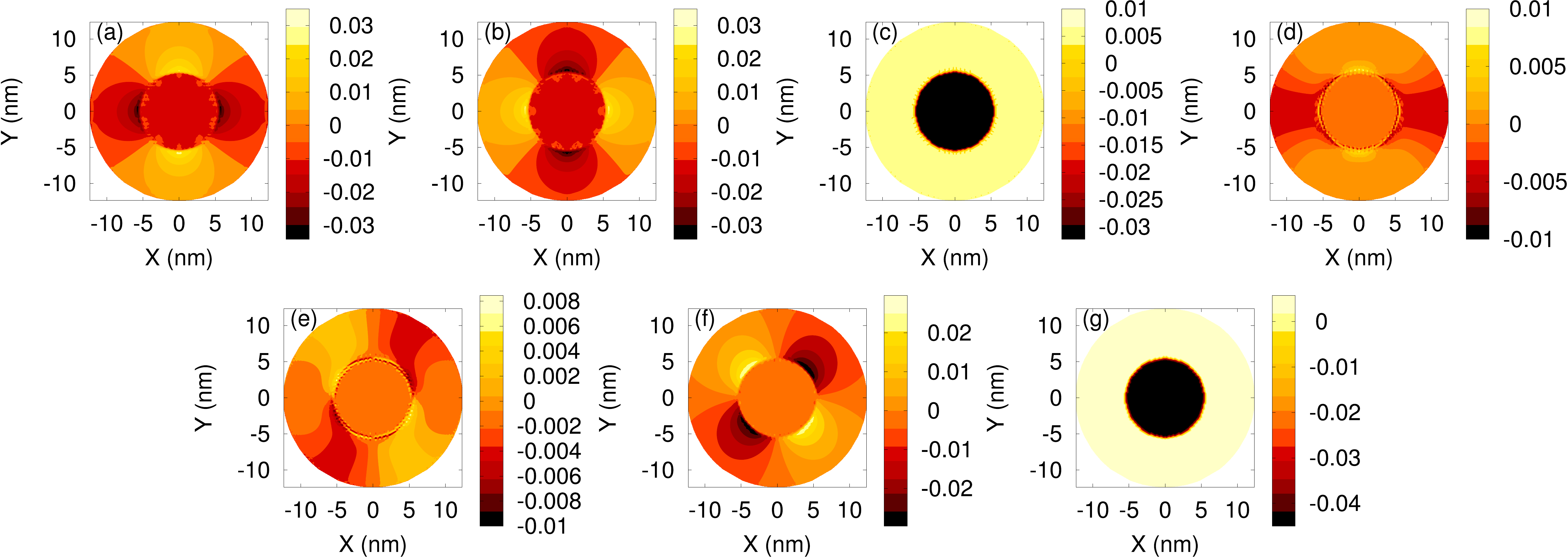}
  \caption{\label{fig:strain_plots_last}\label{fig:zb_cyl_strain}
    Strain in a unit cell of a nanowire with cylindrical core-shell
  geometry. Results obtained from the infinite atomistic model. The
  plots (a)-(g) show the strain components $\varepsilon_{\x \x}$,
  $\varepsilon_{\y \y}$, $\varepsilon_{\z \z}$, $\varepsilon_{\y \z}$,
  $\varepsilon_{\x \z}$, and $\varepsilon_{\x \y}$ and the hydrostatic
  strain $\varepsilon_{\x \x} + \varepsilon_{\y \y} + \varepsilon_{\z
    \z}$, respectively.
}
  
\end{figure*}

Figures \ref{fig:strain_plots_first} to \ref{fig:zb_cyl_strain} show
the strain fields in the $XY$-plane for an infinite wire modeled using
the VFF model and three different core-shell geometries.  The strain
of the hexagonal geometry (depicted in Figs.~\ref{fig:3dWire} and
\ref{fig:CrossSec}) is displayed in \reffig{fig:zb_strain}. All
strains are given with respect to the nanowire oriented coordinate
system ($\x,\y,\z$). Corresponding plots for a parallel hexagonal
geometry, where the core hexagon is rotated by $30^{\circ}$ relative
to that of \reffig{fig:CrossSec}, and for a cylindrical geometry, are
shown in Figs.~\ref{fig:zb_aligned_strain} and
\ref{fig:zb_cyl_strain}. In all cases the nanowire is composed of a
GaAs core and a GaP shell. We discuss the strain components in the
hexagonal geometries, and then compare with the cylindrical geometry.

\subsubsection{Axial Strain}

As discussed in \refsect{sec:PseudoMorph}, we see in several of the
figures that the strain is discontinuous at the core-shell
interface. The axial strain $\varepsilon_{\z \z}$ is displayed in
\reffigpart{fig:zb_strain}{(c)}, and we see that it is constant within
core and shell (as was used in the infinite CE model). The core
material (GaAs) has a larger lattice constant than the shell material
(GaP), and the energetically most favorable core-shell nanowire
configuration will have an interatomic spacing that is in between
those of bulk GaAs and bulk GaP. We therefore see an axial compression
($\varepsilon_{\z \z}<0$) of the core and an axial tension
($\varepsilon_{\z \z}>0$) of the shell.

\subsubsection{Hydrostatic Strain}

We see from the hydrostatic strain in \reffigpart{fig:zb_strain}{(g)}
that the core is compressed while the shell is expanded. In the shell,
the magnitude of the hydrostatic strain
[\reffigpart{fig:zb_strain}{(g)}] is similar to the axial strain
[\reffigpart{fig:zb_strain}{(c)}]. Thus in the shell, hydrostatic strain is
dominated by the axial contribution.

\subsubsection{Planar Tensile Strains}

Tensile strains correspond to expansion or compression along the
coordinate axes. In
\reffigpart{fig:zb_strain}{(a)}, $\varepsilon_{\x \x}$, i.e., the
compression in $\x$-direction, is shown.  The structure is compressed
in the $\x$ direction everywhere along the $\x$-axis. The core is
compressed, as it has to fit into a space too small for it to be at
equilibrium, and the shell is compressed along this axis because it is
pushed outwards by the core. Along the $\y$-axis, the core is again
everywhere compressed, but the shell has been expanded relative to its
equilibrium state. As discussed for the axial strain, the equilibrium
configuration will, parallel to an interface, have an interatomic
spacing in between those the two materials take in bulk. Therefore,
the shell, with a smaller bulk lattice constant, is expanded. A trace
of the same effect can be seen in the $\y$ component of the strain,
\reffig{fig:zb_strain}{(b)}, near the two corners of the core with
$\y=0$. In this case, there is no interface exactly parallel to the
$\y$-axis. Nevertheless, the effect is still visible at the corners.
In \reffigpart{fig:zb_aligned_strain}{(b)}, the core has been rotated
by $30^{\circ}$ relative to the geometry of
\reffigpart{fig:zb_strain}{(b)}. In this case, there is an edge of the
core-shell interface parallel to the $\y$ axis, and we see an
expansion in the shell close to that edge.

\subsubsection{In-plane Shear Strain}

The shear strains $\varepsilon_{\x \y}$, $\varepsilon_{\x \z}$ and
$\varepsilon_{\y \z}$ are nonzero if the local deformation changes the
angles between the basis vectors. Considering the lower left part of
\reffigpart{fig:zb_strain}{(f)}, and applying the same argument we
used for the strain components $\varepsilon_{\x \x}$ and
$\varepsilon_{\y \y}$, we expect radial compression and
circumferential extension of the shell, and we see a nonzero planar
shear strain $\varepsilon_{\x \y}$. The negative value corresponds to
the fact that the right angle between the unit vectors
$\mathbf{\hat{X}}$ and $\mathbf{\hat{Y}}$ is locally enlarged by the
deformation\cite{Nye85}.  Analogously, in the upper left part of the
same figure, the angle between $\mathbf{\hat{X}}$ and
$\mathbf{\hat{Y}}$ is locally shrunk, and this corresponds to a
positive value for $\varepsilon_{\x \y}$.

\subsubsection{Warp Effect}

 The nonzero values for $\varepsilon_{\y \z}$ and $\varepsilon_{\x
   \z}$, shown in \reffigpart{fig:zb_strain}{(d)} and
 \reffigpart{fig:zb_strain}{(e)}, correspond to the fact that a cross
 section in the $\x \y$-plane  is warped
 into a non-flat surface in the strained wire. In other words, when
 the core pushes outwards on the shell, it is energetically favorable
 for the shell to respond by not only becoming compressed radially,
 but also to deform by warping out of the $\x \y$-plane.

\subsubsection{Cylindrical Geometry}

 \reffig{fig:zb_cyl_strain} shows numerical results for a cylindrical
 nanowire geometry. The general features are similar to the
 previously described geometries. The symmetry of the strain
 due to the zinc blende lattice can be seen more clearly in the cylindrical geometry. Figures
 \ref{fig:zb_cyl_strain}(a) and (b) show that $\varepsilon_{\x \x}$
 and $\varepsilon_{\y \y}$ are almost identical in the shell, except
 for the sign. In a significant part of the shell, they are larger in magnitude than both the axial
 [\reffigpart{fig:zb_cyl_strain}{(c)}] and the hydrostatic
 [\reffigpart{fig:zb_cyl_strain}{(f)}] strains. 
 We note also that the warp strains, shown in
 \reffigpart{fig:zb_cyl_strain}{(d)} and
 \reffigpart{fig:zb_cyl_strain}{(e)}, do not vanish in the cylindrical wire. Thus, the warp effect is not solely an artefact of the
 hexagonal nanowire geometry, but an effect of the zinc blende lattice
 structure.

\subsection{Linescans}

\begin{figure}
  \includegraphics[width=7cm]{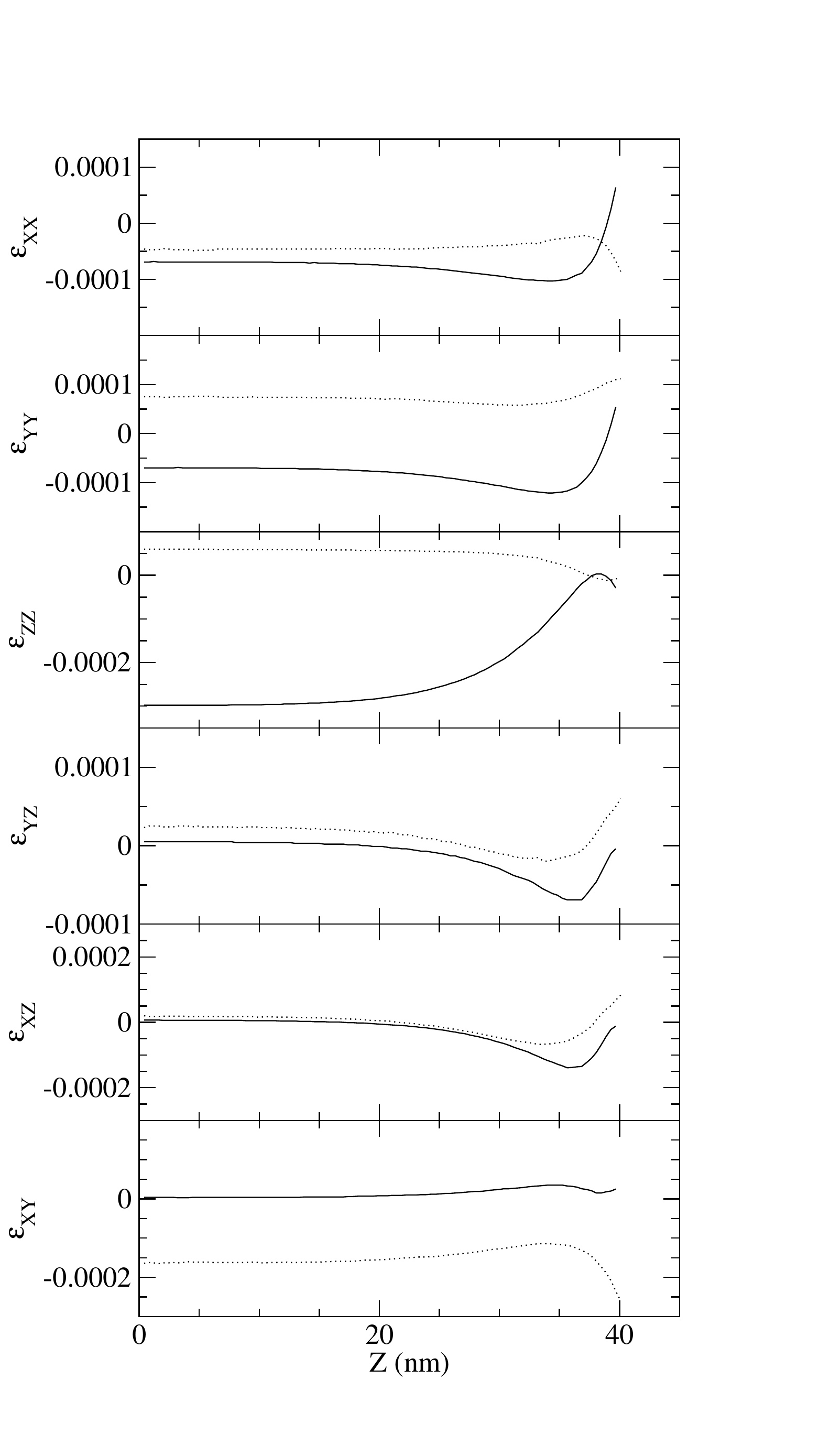}
  \caption{\label{fig:strain_of_z}Linescans of the strain along the nanowire axis. The full-drawn
  lines correspond to $(\x, \y) = (2 \, \mathrm{nm}, 1 \,
  \mathrm{nm})$ (located in the core) and the dotted lines to $(\x,
  \y) = (8 \, \mathrm{nm}, 4 \, \mathrm{nm})$ (in the shell). Results
  obtained from 3D continuum model.}
\end{figure}

\begin{figure}
  \includegraphics[width=7cm]{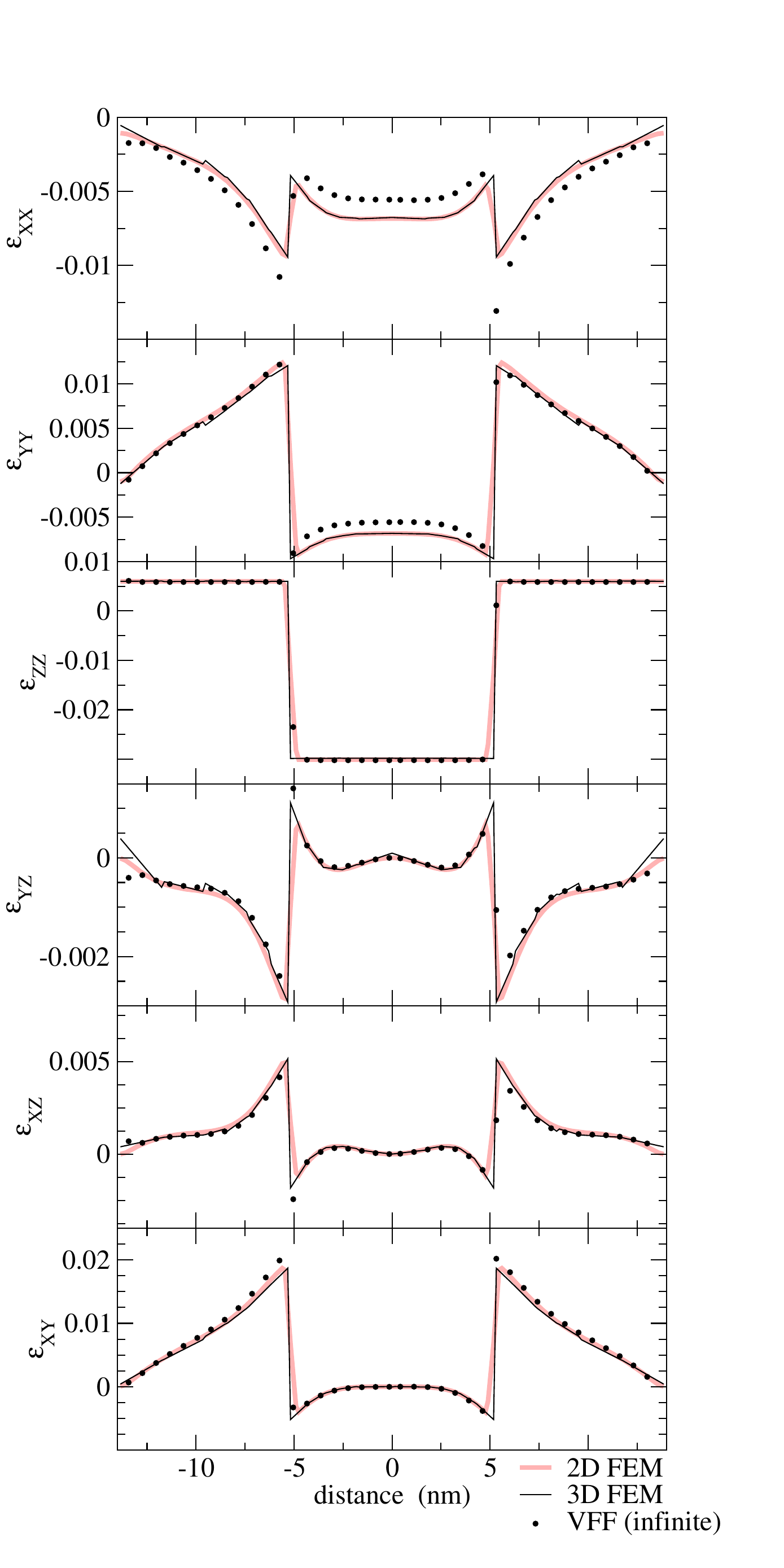}
  \caption{\label{fig:strain_path_1}Linescan along path A as defined in
  \reffig{fig:CrossSec}. Results from three of the four models used
  are shown. The $x$-axis indicates position along the path.
}
  
\end{figure}

\begin{figure}
  \includegraphics[width=7cm]{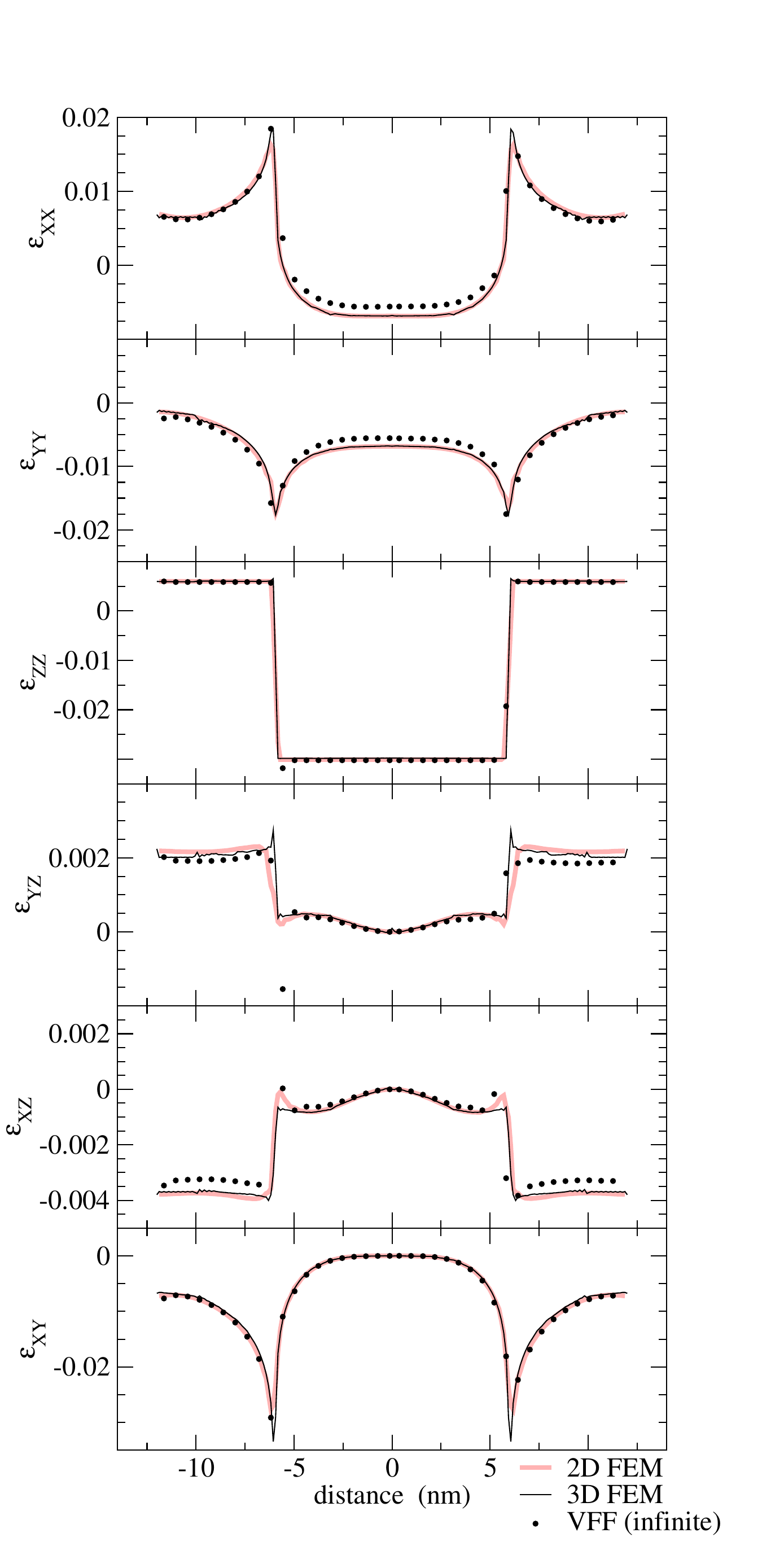}
  \caption{\label{fig:strain_path_6}Linescan along path B as defined in
  \reffig{fig:CrossSec}.  Results from three of the four models used
  are shown. The $x$-axis indicates position along the path.
}
\end{figure}

In \reffig{fig:strain_of_z}, strain components are plotted along paths
which are parallel to the nanowire axis. The solid line corresponds to a path in
the core [$(\x, \y) = (2 \, \mathrm{nm}, 1 \, \mathrm{nm})$], and the dotted line corresponds to a path in the shell [$(\x, \y) = (8 \, \mathrm{nm}, 4 \, \mathrm{nm})$].

As discussed previously, far away from the free ends, the strain 
converges towards its value in an infinite nanowire. We
find that, for points that lie more than 30 nm away from the ends of our
finite nanowire, no strain component deviates by more than $5 \cdot
10^{-6}$ from its value 
at the center of the finite nanowire ($\z=0$). We
interpret this as an indication that our nanowire geometry has sufficient
length and that the strain field at $\z=0$ will correspond to the
strain field in the infinite model.

The behavior of the strain as a function of $\z$ can be
understood as a combination of two effects. The first effect is due to
the hexagonal core-shell geometry and the zinc blende lattice structure and was discussed above. The
second, the Poisson effect,  dominates the strain in the vicinity of the free
ends. The core of the wire is radially compressed by the shell, and
responds by expanding significantly along the axial direction. Far
from the ends, this expansion is prevented by the shell, and
the core is axially compressed, as seen in the plot of
$\varepsilon_{\z \z}$ in \reffig{fig:strain_of_z}. Close to the free
ends, the core bulges out of the wire, giving the end-surfaces of the
nanowire slightly convex shapes. This allows the core to relax more,
as is seen in the same plot, where  the magnitude of $\varepsilon_{\z \z}$ decreases towards
the end of the wire.  The Poisson effect is responsible for the main features of
the $\z$-dependence of the other strain components as well.

Figures \ref{fig:strain_path_1} and \ref{fig:strain_path_6} show a
comparison between the strains of the different models along paths in
an $\x \y$-plane.  The model not shown is the VFF model for a finite
wire. Far from the ends, this model agrees well with the infinite VFF
model.  We observe that the results from the continuum models for
finite and infinite wires are very similar.

Deviations between the VFF and CE models at the core-shell interfaces
are expected. These deviations are unphysical, in the sense that
macroscopic strain is a locally defined phenomenon within a
homogeneous material.  We also see significant systematic deviations
between the VFF and CE models inside the core, and we attribute these
to the discrepancy in effective elastic constants of the VFF and CE
models. As a test of this, we compared calculations from the CE and
VFF models, where we tuned the parameters of the CE model so that the
two models used the same effective elastic constants. A comparison of
the strain fields from those calculations showed good agreement
between the VFF and CE models (comparable to the agreement between our
finite and infinite models).

\section{\label{sec:conclusions}Conclusions}

We have given a self-contained comprehensive account of our
computations of strain distributions in finite and infinite core-shell
nanowires with lattice mismatch using both continuum-elasticity theory
and an atomistic VFF model.  The strain profiles were shown in
Figs.~\ref{fig:strain_plots_first} to \ref{fig:strain_path_6} and were
discussed. The atomistic VFF model has in this work been formulated
within a unit cell of a nanowire with periodic boundary conditions in
the axial direction.  A corresponding continuum elasticity model for
an infinite wire was also defined in a unit cell.  Finally, a long and
finite wire was studied using the continuum-elasticity theory and the
VFF model.

We observe bulging effects at the free ends of the finite wires and
find good agreement between results at a cross section of a finite
wire far away from its ends and at a cross-section of an infinite
wire. The strain distributions obtained in the calculations using the
continuum and atomistic models show qualitatively good
agreement. Quantitatively, deviations remain, but their origin was
discussed. In contrast to the hydrostatic strain, which has a very
simple structure, the individual strain components have very complex
structure.  The non-planar shear strains are generally smaller than
the diagonal components, but non-vanishing. The strain in the core is
dominated by the axial strain, but at the core-shell interface and
just outside it, a richer structure is found in the strain
distributions.

Future work will consist of computing properties such as electronic
structure and phonons based on the results presented here. In those
calculations, extensions to multi-shell structures and
heterostructures combining zinc blende and wurtzite crystals will be considered.

\begin{acknowledgments}
  N.S. and J.G. thank the Swedish Research Council for financial
  support. F.B acknowledges the financial support from the Academy of
  Finland.
\end{acknowledgments}


\begin{thebibliography}{33}
\bibitem{footFBSymm}
{Another 1/6 of the wire is
  obtained through a reflection of the simulated 1/6, and the
  remaining two thirds of the geometry are then obtained as rotated
  copies of the previously obtained third of the wire.}
\end{thebibliography}

\end{document}